\documentclass[a4paper,notitlepage,twocolumn,footnoteinbib,superscriptaddress,longbibliography,10pt]{revtex4-1}

\usepackage{amsmath,bbold,amssymb,amsthm,mathtools,color,listings,graphicx}
\usepackage[colorlinks=true,linkcolor=blue,citecolor=blue]{hyperref}

\newcommand{\rev}[1]{\textcolor{black}{#1}}

\begin{document}

\title{All-optical \rev{scalable} spatial coherent Ising machine}
\author{Marcello Calvanese Strinati}
\email{marcello.calvanesestrinati@gmail.com}
\affiliation{Physics Department, Sapienza University of Rome, 00185 Rome, Italy}
\author{Davide Pierangeli}
\affiliation{Physics Department, Sapienza University of Rome, 00185 Rome, Italy}
\affiliation{Institute for Complex Systems, National Research Council (ISC-CNR), 00185 Rome, Italy}
\author{Claudio Conti}
\affiliation{Physics Department, Sapienza University of Rome, 00185 Rome, Italy}
\affiliation{Institute for Complex Systems, National Research Council (ISC-CNR), 00185 Rome, Italy}
\affiliation{Centro Ricerche Enrico Fermi (CREF), Via Panisperna 89a, 00184 Rome, Italy}
\date{\today}

\begin{abstract}
Networks of optical oscillators simulating coupled Ising spins have been recently proposed as a heuristic platform to solve hard optimization problems. These networks, called coherent Ising machines (CIMs), exploit the fact that the collective nonlinear dynamics of coupled oscillators can drive the system close to the global minimum of the classical Ising Hamiltonian, encoded in the coupling matrix of the network. To date, realizations of large-scale CIMs have been demonstrated using hybrid optical-electronic setups, where optical oscillators simulating different spins are subject to electronic feedback mechanisms emulating their mutual interaction. While the optical evolution ensures an ultrafast computation, the electronic coupling represents a bottleneck that causes the computational time to severely depend on the system size. Here, we propose an all-optical \rev{scalable} CIM with fully-programmable coupling. Our setup consists of an optical parametric amplifier with a spatial light modulator (SLM) within the parametric cavity. The spin variables are encoded in the binary phases of the optical wavefront of the signal beam at different spatial points, defined by the pixels of the SLM. We first discuss how different coupling topologies can be achieved by different configurations of the SLM, and then benchmark our setup with a numerical simulation that mimics the dynamics of the proposed machine. In our proposal, both the spin dynamics and the coupling are fully performed in parallel, paving the way towards the realization of size-independent ultrafast optical hardware for large-scale computation purposes.
\end{abstract}

\maketitle

\section{Introduction}
\label{sec:introduction}
Solving large-scale optimization problems is extremely useful to several different fields of modern science, \rev{with applications ranging} from biology to finance and social science~\cite{Hopfield2554,gilli2011numerical,zhanglifecience,naturecombinatorial,ohzechiarticifialintellingence}. These problems often belong to the non-deterministic polynomial (NP-hard) computational complexity class~\cite{karpcomputercomputation}: Finding the optimal solution requires computational resources that scale exponentially with the size of the system, making these problems intractable using conventional computer architectures. A tremendous amount of interest has been recently attracted by the development of unconventional computational methods (heuristic solvers) to solve probabilistically, but efficiently, large-scale optimization problems. A key observation behind these heuristic methods is that optimization problems can be mapped onto specific classical Ising models efficiently~\cite{10.3389/fphy.2014.00005}, i.e., in a polynomial (P) time. Solving the specific optimization problem then translates into the NP-hard problem of finding the ground state (GS) of the corresponding Ising Hamiltonian~\cite{Barahona_1982}.

In recent years, several physical systems have been demonstrated to evolve according to the classical Ising Hamiltonian, therefore providing valuable \textit{ad hoc} platforms to solve the Ising model for large-scale optimization purposes. Remarkable examples include two-component Bose-Einstein condensates~\cite{Byrnes_2011,yamamotoboseeinstein2013}, superconducting circuits~\cite{Johnson2011}, trapped ions~\cite{monroequantumsimulation2010,bollingerengineering2012}, digital computers~\cite{king2018emulating,Tiunov:19,hgoto2019cim,hgoto2021cim}, electrical oscillators~\cite{chou2019}, optoelectronical oscillators~\cite{bohm2019}, polariton condensates~\cite{kalininpolariton2017,PhysRevLett.121.235302,kalinin2019}, laser networks~\cite{Utsunomiya2019,gershenzonrapidlaser2019}, and coupled optical parametric oscillators (OPOs)~\cite{PhysRevA.88.063853,marandi2014cim,takata2016cubic,nphoton.2016.68,hamerlyfristratedchain2016,PhysRevA.96.043850,Wang2017,Inagaki603,1805.05217,PhysRevLett.122.213902,PhysRevLett.123.083901,10.1007/978-3-030-19311-9_19,gaeta2020,Pierangeli:20}, which are the focus of this work. These networks, called coherent Ising machines (CIMs) exploit the fact that, when driven above the oscillation threshold, a second-order phase transition takes place~\cite{goto1959,landauer1971}: In the long-time limit, the phase of each OPO takes values $0$ or $\pi$ with respect to the reference phase enforced by the pump. Because of the bistable nature of its phase, an OPO is suitable to simulate a classical Ising spin, and systems of coupled OPOs in proper \rev{conditions} can simulate the dynamics of coupled Ising spins~\cite{hamerlyfristratedchain2016,PhysRevLett.126.143901,arXiv:2105.07591}.

Nowadays, major issues in realizing CIMs for realistic optimization problems involve, on one hand, the physical conditions (e.g., temperature) in which the machine has to operate, and on the other hand, the scalability and the connectivity that these systems can implement. In this respect, photonic systems offer a versatile platform to realize large-scale CIMs with general connectivity, while working at room temperature and being constructed from off-the-shelf components. \rev{An implementation of an all-optical CIM with few spins using time-multiplexed OPOs has been reported in Refs.~\cite{marandi2014cim,takata2016cubic}, where different OPOs are different temporal pulses within a nonlinear cavity, and optical delay lines are used to couple different OPOs. This approach allows the realization of arbitrary coupling topology, but cannot be scaled up to a large number of spins. An all-optical CIM with a large-number of spins was reported in Ref.~\cite{nphoton.2016.68}, implementing the one-dimensional nearest-neighbor coupling via a Mach-Zehnder interferometer. While this other approach allows the implementation of several spins, the coupling topology was limited to nearest-neighbor coupling.} \rev{Large-scale CIMs with arbitrary coupling topology have} been demonstrated using \rev{time-multiplexed OPOs in} hybrid optical-electronic systems~\cite{Haribara2016,Inagaki603}, or optoelectronic oscillators~\cite{bohm2019}: Optical signals evolve in time subject to electronic measure and feedback mechanisms to emulate the spin-spin interaction. The optical nature of the setup ensures ultrafast computation. However, the presence of the electronic feedback inherently represents a ``bottleneck'' that introduces an additional computational time that scales quadratically with the system size~\cite{PhysRevApplied.15.034087}. The realization of a scalable fully optical CIM without the hindrance of electronic components is thus highly desirable.

\begin{figure}[t]
\centering
\includegraphics[width=8.7cm]{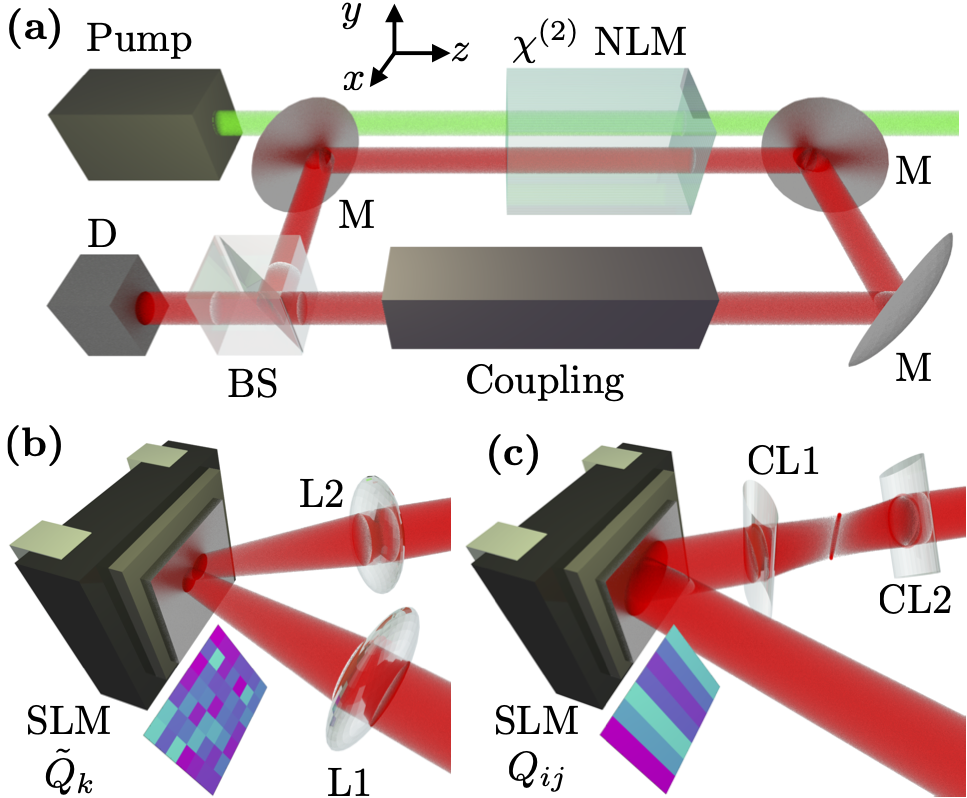}
\caption{\textbf{(a)} All-optical spatial CIM. The setup consists of the following: (i) $\chi^{(2)}$ NLM pumped by a laser at wavelength $\lambda_p$ non-resonant with the cavity mirrors (M), which amplifies the signal at $\lambda_s=2\lambda_p$, (ii) Coupling mechanism, (iii) Extraction of the signal by a beam splitter (BS) with reflection and transmission coefficients $R_{\rm out}$ and $T_{\rm out}=\sqrt{1-R_{\rm out}^2}$, respectively, and detection (D). The coupling is implemented using different configurations of the SLM depending on the implemented graph. \textbf{(b)} Circulant graphs: A first lens (L1) transforms the signal in Fourier space, the SLM multiplies the FT of the field by $\tilde Q_k$, and a second lens (L2) transforms the modulated field back to real space. \textbf{(c)} General graphs, based on the vector-matrix multiplication scheme. The SLM works in real space, multiplying the incoming field by $Q_{ij}$. The modulated signal is focused along the $y$-axis by a cylindrical lens (CL1), defocused, and rotated by $90^\circ$ on the $xy$-plane by another cylindrical lens (CL2) with $45^\circ$ rotated optical axis. The pixels in panels \textbf{(b)} and \textbf{(c)} depict the OPOs encoded on the $xy$-plane: An amplitude $A_{j,\tau}$ is defined as a single pixel in panel \textbf{(b)}, whereas as a column of $M_y$ pixels in panel $\textbf{(c)}$.}
\label{fig:schemeoftheproposedexperiment}
\end{figure} 

A step towards the realization of an all-optical \rev{scalable} CIM has been recently made by exploiting spatial degrees of freedom of light~\cite{PhysRevLett.122.213902,huang2020spatial}. This approach relies on the usage of spatial light modulators (SLM\rev{s}) to encode a spin variable $\sigma_j$ into the binary phase modulation of the optical field shining the $j$-th pixel of the SLM. An electronic feedback mechanism was also present. To date, a proposal of an all-optical scalable CIM implementing an arbitrary coupling topology is still missing. In this \rev{paper, we} propose and theoretically validate an all-optical \rev{scalable} spatial CIM with fully-programmable coupling. \rev{We first propose two different configurations of the SLM to realize two different classes of coupling, and then estimate the computational performance of our machine by means of a numerical simulation that closely mimics the temporal dynamics of coupled OPOs within a parametric cavity. We find that our proposed setup converges close to the minimum of the Ising Hamiltonian after a computational time that is orders of magnitude smaller compared to existing hybrid electro-optical setups.}

\rev{This paper is organized as follows. In Sec.~\ref{sec:schemeofthesetup}, we discuss the scheme of our proposed setup. In Sec.~\ref{sec:numericalresults}, we show our numerical results, and draw our conclusions in Sec.~\ref{sec:conclusions}.}

\section{Scheme of the proposed setup}
\label{sec:schemeofthesetup}
The scheme of our machine is shown in Fig.~\ref{fig:schemeoftheproposedexperiment}\textbf{a}. We consider an optical parametric cavity with a second-order nonlinear medium ($\chi^{(2)}$ NLM) of length $L$, shone by a pump laser (green beam) at wavelength $\lambda_p$. We take $z$ as the direction of propagation of light. The pump beam within the NLM parametrically amplifies degenerate pairs of signal and idler photons (red beam) at wavelength $\lambda_s=2\lambda_p$ via spontaneous parametric down conversion. The spatial configuration of the signal wavefront on the $xy$-plane encodes the amplitudes $\{A_{j,\tau}\}$ of the $N$ OPO fields ($j=1,\ldots,N$) at round trip number $\tau$. These amplitudes are in general complex, i.e., the phase of the optical field on the $xy$-plane can take any value. However, the presence of the $\chi^{(2)}$ NLM forces the optical phases to be either $0$ or $\pi$ with respect to the pump phase, making the amplitudes $\{A_{j,\tau}\}$ effectively real, as detailed below.

Different configurations of the SLM with $M_x\times M_y$ pixels realize different couplings between different OPOs, as shown in panels \textbf{(b)} and \textbf{(c)}. In panel \textbf{(b)}, the SLM is placed at the focal plane of a first lens (L1). The discretization of the field in real (and thus momentum) space is enforced by the SLM: Different pixels define different OPO amplitudes, thereby defining $N=M_x\times M_y$ different OPOs. Since in commercial SLMs one typically has $M_x,M_y\sim10^3$, this scheme allows to define $N\sim10^6$ OPOs. The SLM acts as a programmable matrix with a transmission function $\tilde Q_k$, which multiplies at each round trip the Fourier transform (FT) of the incoming OPO amplitudes $\tilde A_{k,\tau}={\rm FT}[\{A_{j,\tau}\}]$. A second lens (L2) with same focal length as L1 transforms the modulated fields $\tilde Q_k\tilde A_{k,\tau}$ back to real space, yielding the inverse Fourier transform (IFT) of $\tilde Q_k\tilde A_{k,\tau}$. By convolution theorem, the resulting field on each pixel after the FT-SLM-IFT sequence is $A'_{j,\tau}=\sum_iQ_{i-j+1}A_{i,\tau}$, where $Q_j$ is the IFT of $\tilde Q_k$. The output takes the form of a coupled field $A'_{j,\tau}=\sum_iQ_{ij}A_{i,\tau}$, where the coupling matrix $\mathbf{Q}$ has entries $Q_{ij}\equiv Q_{i-j+1}$. This matrix represent a rotationally invariant (or circulant) graph~\cite{davis1994circulant}, where all nodes are equivalent. Notable examples are the nearest-neighbor Ising chain and the M\"obious ladder~\cite{marandi2014cim,takata2016cubic,nphoton.2016.68,1805.05217}.

While the coupling scheme in panel \textbf{(b)} gives the possibility to encode a large number of OPOs, granting a straightforward experimental implementation, it allows the implementation of a limited class of graph. To overcome this issue, we propose in panel \textbf{(c)} a different scheme for a general coupling matrix $\mathbf{Q}$. This setup is based on the vector-matrix multiplication scheme~\cite{Spall:20}: The different OPOs are arranged on the $xy$-plane as different column vectors, such that the signal wavefront shines all $M_y$ pixels on a given column of the SLM with a uniform field. The SLM multiples in real space the vectorized signal with amplitude $A_{j,\tau}$ by $Q_{ij}$, such that the amplitude at the point $(i,j)$ of the field after the SLM is $Q_{ij}A_{j,\tau}$. A cylindrical lens (CL1) focuses the signal wavefront onto a single column along the $y$-axis, whose amplitude at point $i$ is given by $A'_{i,\tau}=\sum_jQ_{ij}A_{j,\tau}$. Propagation in free space defocuses the signal on the $xy$-plane, obtaining a vectorized signal arranged as a row matrix. Subsequent rotation by $90^\circ$ of the field by a second cylindrical lens (CL2) recovers the structure of the signal as column vectors. Now, each column encodes the amplitude $A'_{j,\tau}=\sum_iQ_{ij}A_{i,\tau}$, i.e., a coupled field with general $Q_{ij}$. As such, this scheme implements any coupling matrix, but with the drawback that the OPOs  need to be redundantly defined over $M_y$ pixels, limiting to $N=M_x$ the number of OPOs in the system.

We stress that the two coupling schemes presented here are fully optical and process all interactions in parallel, without need of electronic feedbacks. Since in our scheme the propagation of the OPOs within the cavity also occurs in parallel, our scheme realizes a size-independent large-scale spatial CIM~\cite{PhysRevApplied.15.034087}, with critical advantages in terms of scaling and computational time compared to the existing hybrid optical-electronic devices. 

In our setup, the binary nature of the phase of each OPO is enforced by the $\chi^{(2)}$ NLM in Fig.~\ref{fig:schemeoftheproposedexperiment}\textbf{a}. We inject into the NLM a pump field with spatially uniform wavefront that, at each pixel $j$ on the $xy$-plane, mixes inside the NLM with the signal field. The subsequent dynamics along the $z$-axis, independently at each point, follows the second-order nonlinear wave equation~\cite{boyd2008nonlinear} for the degenerate signal field $\{A_{j,\tau}\}$ and pump $\{B_{j,\tau}\}$:
\begin{equation}
\frac{dB_{j,\tau}}{dz}=-\kappa\,A_{j,\tau}^2 \qquad \frac{dA_{j,\tau}}{dz}=\kappa\,B_{j,\tau}A^*_{j,\tau} \,\, ,
\label{eq:secondorderwaveequation1}
\end{equation}
where the star denotes complex conjugation. Here, $\kappa=2\pi\chi^{(2)}/\lambda_s n^2$, where $\chi^{(2)}$ and $n=n(\lambda_s)$ are the nonlinear coefficient and the index of refraction of the NLM, respectively, and $\lambda_s=2\pi/k_s$. We assume perfect phase matching $2k_s=k_p$, where $k_{s(p)}$ is the wave vector of the signal (pump), which is achieved by tuning the temperature of the NLM to equalize the index of refraction for the signal and pump. To show phase-dependent amplification, we rewrite by omitting $j$ and $\tau$ in Eq.~\eqref{eq:secondorderwaveequation1} $A(z)=u(z)\,e^{i\phi(z)}$ and $B(z)=u_p(z)\,e^{i\phi_p(z)}$, where $u$ ($u_p$) and $\phi$ ($\phi_p$) are the amplitude and phase of the signal (pump), respectively. This allows to separate the dynamics of the relative phase $\theta=\phi_p-2\phi$ and of the amplitudes $u$ and $u_p$ in Eq.~\eqref{eq:secondorderwaveequation1} as
\begin{eqnarray}
\frac{du}{d(\kappa z)}&=&uu_p\cos(\theta) \qquad \frac{du_p}{d(\kappa z)}=-u^2\cos(\theta) \nonumber\\
\frac{d\theta}{d(\kappa z)}&=&\frac{\sin(\theta)}{u_p}\left(u^2-2u^2_p\right) \,\, .
\label{eq:secondorderwaveequation2}
\end{eqnarray}
From Eq.~\eqref{eq:secondorderwaveequation2}, one can see that the evolution of $\theta$ has two fixed points (modulo $2\pi$): $\theta=0$, i.e., $\phi-\phi_p/2=0,\pi$ and $\theta=\pi$, i.e., $\phi-\phi_p/2=\pm\pi/2$, corresponding to two distinct regimes: (i) Parametric amplification, where energy is transferred from the pump to the field, and (ii) Up-conversion, where energy is converted from the signal to the pump. Focusing on the first case, which is our case of interest, $\theta$ flows towards $\theta=0$, fixing $\phi$ to be either $0$ or $\pi$ with respect to $\phi_p/2$, thereby manifesting phase-dependent amplification. In terms of the original variables, taking $\phi_p=0$ (real pump), the evolution along $z$ amplifies the real part of the fields ${\rm Re}[A_{j,\tau}]$ and suppresses their imaginary parts ${\rm Im}[A_{j,\tau}]$.

\begin{figure}[t]
\centering
\includegraphics[width=4.24cm]{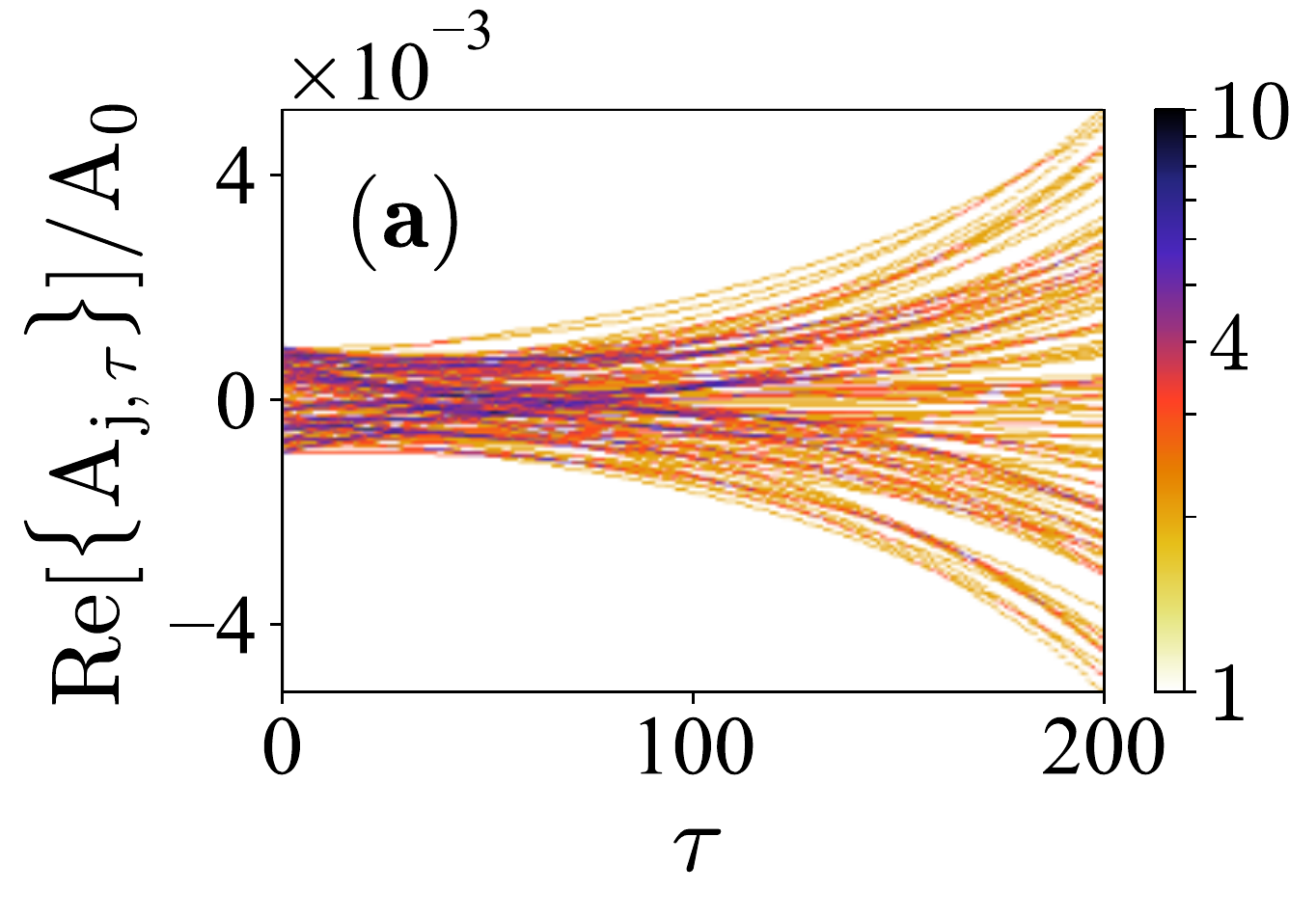}
\includegraphics[width=4.31cm]{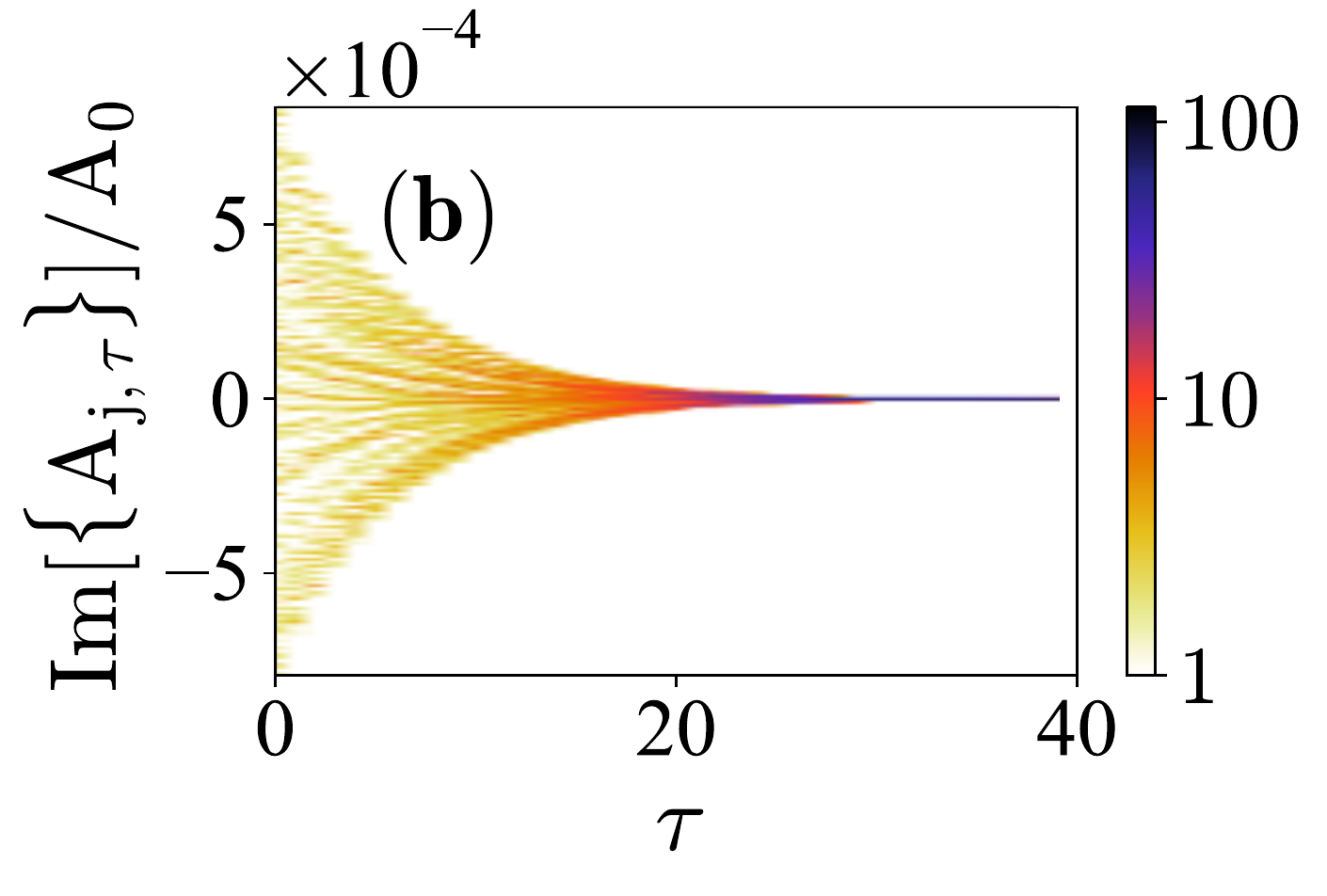}
\caption{Histogram of time evolution of the \textbf{(a)} real and \textbf{(b)} imaginary part of the $N=112$ OPO amplitudes (in units of $A_0$, see text), computed using Eq.~\eqref{eq:nonlinearmap1}, for short times. As evident, the cavity dynamics enhances the real part of the OPO amplitudes and suppresses their imaginary part.}
\label{fig:cavitydynamics1}
\end{figure}

\begin{figure*}[t]
\includegraphics[width=18cm]{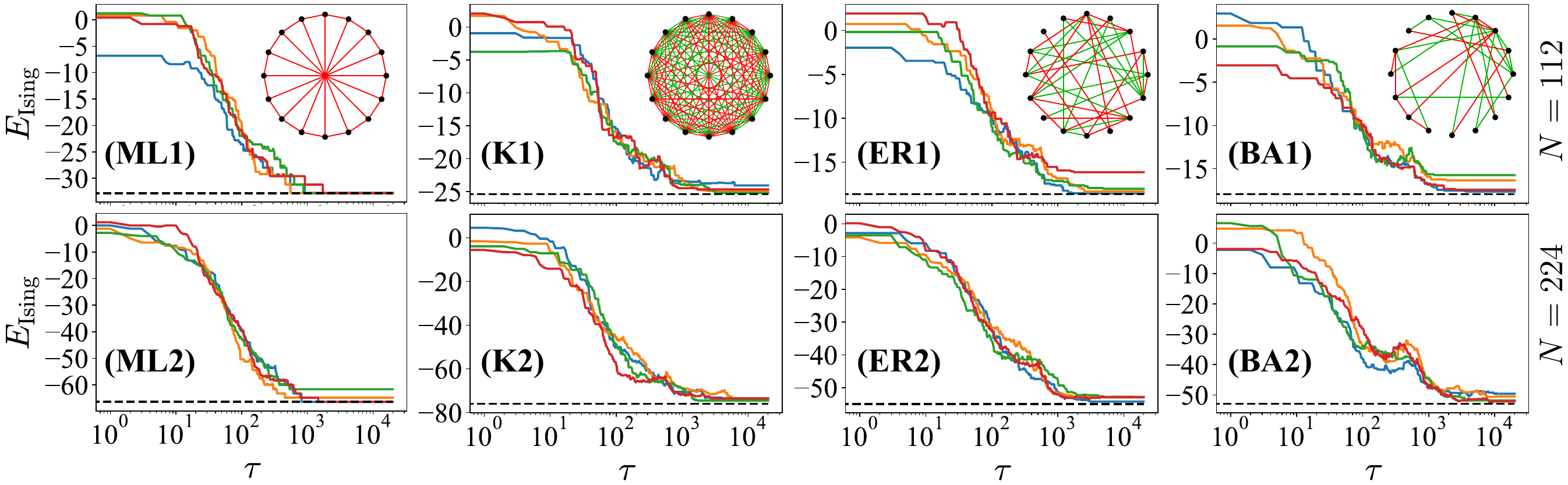}
\caption{\rev{Ising energy from the OPO phases during the round trips for \textbf{(ML1),(ML2)} M\"obius ladder, \textbf{(K1),(K2)} complete graph, \textbf{(ER1),(ER2)} Erd\H{o}s-R\'enyi graph, and \textbf{(BA1),(BA2)} Barab\'asi-Albert graph. Upper panels refer to $N=112$, while lower panels to $N=224$. Different lines refer to different initial conditions. Horizontal black dashed lines mark the GS energy found by diagonalizing $\mathbf{J}$ for ML, and the minimal energy found by Metropolis annealing for K, ER, and BA. The insets in the upper panels depict the different connectivities in the four cases (black dots are nodes, and red and green lines depict negative and positive edge weights, respectively), where $N=16$ is used for illustration purposes only.}}
\label{fig:cavitydynamics2}
\end{figure*}

\section{Numerical results}
\label{sec:numericalresults}
We now discuss the experimental realizability of our setup in Fig.~\ref{fig:schemeoftheproposedexperiment}. We follow Refs.~\cite{Calvanese_Strinati_2020,PhysRevLett.126.143901} using realistic experimental values of the system parameters. Each OPO amplitude $A_{j,\tau}$ evolves into $A_{j,\tau+1}$ after a round trip by undergoing (i) Parametric amplification within the NLM [Eq.~\eqref{eq:secondorderwaveequation1}], (ii) Coupling, and (iii) Measurement and losses. We then capture the OPO dynamics by the following map:
\begin{equation}
A_{j,\tau+1}=R_{\rm out}\sum_lQ_{jl}\,{\rm NLM}[A_{l,\tau}] \,\, .
\label{eq:nonlinearmap1}
\end{equation}
In Eq.~\eqref{eq:nonlinearmap1}, ${\rm NLM}[A_{l,\tau}]$ represents the $l$-th OPO amplitude at the exit of the NLM (i.e., for $z=L$), which is computed by integrating Eq.~\eqref{eq:secondorderwaveequation1} for $z\in[0,L]$ with initial conditions $A_{j,\tau}(z=0)=A_{j,\tau-1}$ and $B_{j,\tau}(z=0)=B_0$, where $B_0$ is the uniform pump amplitude at the entrance of the NLM, which provides the gain. There are two sources of loss: The SLM transmission function $\mathbf{Q}$, and measurement and intrinsic loss encoded in $R_{\rm out}$. The balance between gain and losses during the initial round trips defines the value $B_{0,\rm th}$ of the oscillation threshold: $B_{0,\rm th}=-\log[R_{\rm out}\rho(\mathbf{Q})]/\kappa L$, where $\rho(\mathbf{Q})$ is the spectral radius of $\mathbf{Q}$.

We shine the NLM with a pump laser at $\lambda_p=\rev{532}\,{\rm nm}$ with spot radius $w_0=100\,{\rm \mu m}$. The physical properties of the NLM are encoded in $\chi^{(2)}$, $\lambda_s$, and $n$ (determining $\kappa$), and $L$. Here, we use $\chi^{(2)}=10^{-11}\,{\rm m/V}$, $\lambda_s=1064\,{\rm nm}$, $n=2$ ($\kappa\simeq1.48\times10^{-5}\,{\rm V^{-1}}$), and $L=10\,{\rm cm}$, which defines the characteristic length scale in our simulations. To integrate Eq.~\eqref{eq:secondorderwaveequation1}, we use as characteristic field amplitude $A_0\simeq6.77\times10^3\,{\rm V/m}$, which yields the numerical rescaled nonlinear constant $\tilde\kappa\coloneqq\kappa z_0A_0=10^{-2}$, and thus integrate Eq.~\eqref{eq:secondorderwaveequation1} for $z\in[0,1]$ (in units of $L$). At $\tau=0$, the signal consists of white noise with amplitude $10^{-3}\,A_0$.

The SLM transmission function $\mathbf{Q}$ is written by including self-interaction and off-diagonal coupling terms $\mathbf{J}$, which is a real symmetric matrix: $\mathbf{Q}=a\mathbb{1}+b\mathbf{J}$, where $\mathbb{1}$ is the identity matrix. Since the SLM provides phase and amplitude modulation, $Q_{ij}$ is in general a complex number with $|Q_{ij}|<1$. The lossy nature of the SLM implies $\rho(\mathbf{Q})<1$. To benchmark our proof-of-principle machine, we simulate $N=112$ coupled OPOs and solve the MAX-CUT problem for four undirected graphs for which the optimization problem belongs to different classes of computational complexity (P and NP-hard)~\cite{kalinin2020,PhysRevLett.126.143901}: The M\"obius ladder (ML)~\cite{harary1967moebius}, which is a circulant P-graph realized with the scheme in Fig.~\ref{fig:schemeoftheproposedexperiment}\textbf{b}, and three NP-hard graphs, specifically the random Erd\H{o}s-R\'enyi (ER) and scale-free Barab\'asi-Albert (BA) graphs~\cite{RevModPhys.74.47} with approximately $20\%$ edge density, and the random complete (K) graph~\cite{schneider1993graphs}, realized with the scheme in Fig.~\ref{fig:schemeoftheproposedexperiment}\textbf{c}. For the ML, the nonzero entries of $\mathbf{J}$ are $J_{i,i+1}=\alpha$ and $J_{i,i+N/2}=\alpha$, with negative $\alpha$. For the ER and BA graphs, $J_{ij}=0,\pm\beta$, where the three values are randomly chosen from the appropriate probability distribution to yield the chosen edge density~\cite{RevModPhys.74.47}, while for the K graph, $J_{ij}=\pm\gamma$, where the sign is randomly chosen with equal probability. We set $a=0.96$, $b=0.04$, $\alpha=-0.2$, and $\beta=0.05$, and $\gamma=0.03$, for which $\rho(\mathbf{Q})\simeq0.98$ in all cases. In this setup, by setting $T_{\rm out}=\sqrt{0.1}$, we obtain threshold pump powers $P_{\rm th}\sim200\,{\rm mW}$.

\begin{table}[t]
\centering
\begin{tabular}{c|c|c|c|c}
\hline\hline
 & \textbf{ML} & \textbf{K} & \textbf{ER} & \textbf{BA} \\\hline
 $\mathbf{N=112}$ & $1.2\,B_{\rm 0,th}$ & $1.33\,B_{\rm 0,th}$ & $1.3\,B_{\rm 0,th}$ & $1.3\,B_{\rm 0,th}$\\\hline
  $\mathbf{N=224}$ & $1.2\,B_{\rm 0,th}$ & $1.36\,B_{\rm 0,th}$ & $1.3\,B_{\rm 0,th}$ & $1.3\,B_{\rm 0,th}$\\\hline\hline
\end{tabular}
\caption{\rev{Values of the pump amplitude $B_0$ at the entrance of the NLM used in the simulation, for the data in Fig.~\ref{fig:cavitydynamics2}.}}
\label{table:dataforisingenergytime}
\end{table}

We first show in Fig.~\ref{fig:cavitydynamics1} the histogram of the time evolution of \textbf{(a)} ${\rm Re}[A_{j,\tau}]$, and \textbf{(b)} ${\rm Im}[A_{j,\tau}]$, for short times, specifically for ML. As evident, ${\rm Re}[A_{j,\tau}]$ is exponentially amplified, whereas ${\rm Im}[A_{j,\tau}]$ is instead suppressed. This confirms that the system is correctly in the phase-dependent amplification regime. Next, we simulate the OPO dynamics for the chosen graphs and compute the Ising energy from OPO phase configuration as $E_{\rm Ising}(\tau)=-(1/2)\sum_{ij}J_{ij}\sigma_i(\tau)\sigma_j(\tau)$, where $\sigma_i(\tau)={\rm sgn}(A_{i,\tau})$ and ${\rm sgn}(\cdot)$ is the sign function. \rev{We compare the energy for two different values of $N$, namely $N=112$ and $N=224$. The result is shown in} Fig.~\ref{fig:cavitydynamics2}. \rev{Different panels refer to different graphs and different $N$ as in the labels, using the values of the pump amplitude $B_0$ at the entrance of the NLM given in Table~\ref{table:dataforisingenergytime}}. Different solid lines refer to different initial conditions of the fields. The black horizontal lines mark the minimal Ising energy: The P nature of the problem with the ML allows to find this value exactly by diagonizaling $\mathbf{J}$, selecting the eigenvector with maximal eigenvalue~\cite{PhysRevLett.126.143901}. Instead, for the NP-hard cases of K, ER, and BA, we estimate the minimal value by a Metropolis annealing algorithm. The fact that the steady-state value of $E_{\rm Ising}$ coincides with the exact minimal value for ML, while sometimes it does not for K, ER, and BA, reflects the different computational complexity of the optimization~\cite{kalinin2020,PhysRevLett.126.143901}.

The key result in Fig.~\ref{fig:cavitydynamics2} is that our system approaches a steady state \rev{with energy close to the minimum energy of the corresponding Ising Hamiltonian} after about $10^3$ round trips, \rev{for both values of $N$}. Since in our scheme all OPOs evolve in time in parallel, our device allows to use short cavities (typically $D\sim1\,{\rm m}$ long) that in turn ensures short round trips times $\tau_{\rm RT}=2nD/c\sim10\,{\rm ns}$, independent of $N$. We then estimate the total computational time as approximately $10\,{\rm \mu s}$. As such, the parallelization of the dynamics allows to envision orders of magnitude shorter computational time compared to existing CIM realizations~\cite{Inagaki603,Haribara_2017,s41534-017-0048-9,1805.05217}.

\section{Conclusions}
\label{sec:conclusions}
In conclusion, we propose a fully-optical scalable spatial CIM implementing different connectivities. The binary nature of the signal phase at different points on the wavefront is enforced by the NLM within the parametric cavity, and the spatial discretization of the wavefront is defined by the SLM within the cavity implementing the spin-spin optical coupling. The number of spins that our system encodes critically depends on the specific configuration of the SLM. To implement a fully-programmable CIM, we propose a setup based on the vector-matrix multiplication scheme, where the SLM works in real space of the field. This scheme allows the implementation of any graph, however limiting the number of spins to $N\sim10^3$ due to the redundant encoding of the spin variables on the SLM pixels. We then propose an alternative coupling scheme with the SLM working in momentum space of the field, which allows the implementation of a limited class of graphs but it can host $N\sim10^6$ spins.

The all-optical nature of our machine presents a step towards the realization of large-scale scalable CIMs. First, in our proposal, both the OPO dynamics and their mutual coupling are fully parallelized, which makes the reach for the optimal solution size independent. Second, the parallel encoding of all OPOs allows for short cavity lengths, and thus drastically smaller computational time, compared to state-of-the art realizations with hybrid electronic-optical setups, where the OPOs are arranged as a temporal sequence of pulses and cavity lengths of approximately $1\,{\rm km}$ are needed. Another important advance of our proposal compared to existing realizations is that no measurement is performed during the time evolution to realize the mutual interaction. This makes our machine suitable to study fundamental features of coupled OPOs beyond optimization, like the emergence of robust macroscopic quantum entanglement~\cite{drummond2021,PhysRevA.104.013715}. 

\section*{Acknowledgements}
We acknowledge funding from Sapienza Ricerca, PRIN PELM (20177PSCKT), QuantERA ERA-NET Co-fund (Grant No. 731473, Project QUOMPLEX), H2020 PhoQus Project (Grant No. 820392).


%

\end{document}